\documentclass[prl,twocolumn]{revtex4}
\usepackage{graphicx}

\begin{document}

\title{Signatures of spin-charge separation in scanning probe microscopy}

\author{Iddo Ussishkin}
\author{Leonid I. Glazman}
\affiliation{William I. Fine Theoretical Physics Institute, University
of Minnesota, Minneapolis, Minnesota 55455}

\date[]{June 30 , 2004}

\begin{abstract}
We analyze the effect of an auxiliary scatterer, such as the potential
of a scanning tip, on the conductance of an interacting one-dimensional
electron system. We find that the differential conductance for
tunneling into the end of a semi-infinite quantum wire reflects the
separation of the elementary excitations into spin and charge modes.
The separation is revealed as a specific pattern in the dependence of
the conductance on bias and on the position of the scatterer.
\end{abstract}

\maketitle

Interaction between electrons has a profound effect on the properties
of one-dimensional electron systems. Basic notions of Fermi liquid
theory are no longer applicable, as is manifested, e.g., in the absence
of well-defined fermionic quasiparticles related to the non-interacting
state. Instead, the low-energy physics of these systems is described by
the concept of a Luttinger liquid~\cite{haldane}.

One consequence of the inter-electron interaction is the separation of
elementary excitations into two branches with different velocities,
corresponding to spin and charge excitations. This spin-charge
separation is apparent already within the random-phase
approximation treatment of the interaction. The relation between the
bare electron degrees of freedom and the elementary excitations of a
Luttinger liquid, manifested in the suppression of the electron density
of states at low energies~\cite{review}, requires more sophisticated
techniques such as bosonization~\cite{haldane}.

The suppression of the density of states was demonstrated in several
tunneling experiments~\cite{bockrath,yao}. In contrast, the separation
into spin and charge modes has proved harder to observe. The simplest
idea, to measure the charge- and spin-density response functions, is
problematic because coupling to the spin mode is difficult. 
Another approach is provided by electron tunneling experiments, either
into a nearby two-dimensional electron gas (2DEG)~\cite{altland,si}, or
into a second one-dimensional system~\cite{ophir,yaroslav,zulicke,balents}.
The latter experiment was recently carried out by Auslaender \emph{et al.} 
in quantum wires grown in a GaAs heterostructure~\cite{ophir,yaroslav}.
In these experiments, the differential conductance 
was measured as a function of voltage and
applied magnetic field. Patterns of oscillations in the measured conductance,
arising from finite size effects, showed modulations which provide evidence
for spin-charge separation in the two wires~\cite{yaroslav}.
Another possibility is to use scanning
tunneling microscopy (STM) to measure the local density of states,
close to the edge of the system~\cite{eggert1}, in a finite
system~\cite{eggert2}, or next to an impurity~\cite{kivelson}.
A difficulty in STM experiments is the need to
maintain a small distance between the wire and the tip, necessary for
electron tunneling. This is possible in carbon nanotubes; in recent
experiments~\cite{lemay}, however, interaction between the electrons
was suppressed by the metallic substrate, and the results were
adequately interpreted in terms of non-interacting electrons.

A scanning tip may also be placed further away from the wire, thus
prohibiting tunneling, but still creating a local potential, which
scatters electrons.  The conductance of the system may then be measured
as a function of tip position.  This scanning probe microscopy (SPM) technique
was successfully used for imaging the flow of
electrons through a quantum point contact in a 2DEG~\cite{topinka}.
Similar techniques were also used to image other systems, including
confined electrons in a 2DEG~\cite{crook}, multichannel one-dimensional
wires~\cite{ihn}, and quantum dots in nanotubes~\cite{woodside}.

In this Letter, we analyze the use of SPM in one dimension.
We consider electrons tunneling into a semi-infinite one-dimensional system 
through a fixed tunnel junction at its end.
The scanning tip is used to create a weak potential scatterer along the wire
at a distance $x_0$ away from its end.
The tunneling conductance is then calculated as a function of tip position $x_0$.
Note the difference from the STM setup (Ref.~\cite{eggert1,kivelson,eggert2})
in which the tunneling occurs at the tip position. 
 
We find that the tunneling conductance exhibits
oscillations as a function of the applied voltage $V$ and the position
$x_0$ of the scatterer. The nature of these oscillations depends on the
strength of the electron-electron interaction. At moderate or weak
interaction, spin-charge separation is readily identifiable as a
beating pattern, which allows one to extract the velocities of spin and
charge modes. Notably, this information can be extracted from the
voltage dependence alone, which may be useful in some experimental
setups with a fixed position of the scatterer. At a strong interaction,
the dominant contribution oscillates only as a function of $x_0$, with
wavevector $2 k_F$ (here $k_F$ is the electron Fermi wavevector).

To analyze the effect of an auxiliary scatterer on the current $I$ of
electrons which tunnel into the end of a one-dimensional wire, we
express the current,
\begin{equation}
I \propto \int d \epsilon \, [ n_F (\epsilon) - n_F (\epsilon + eV) ]
\, \nu_{\text{lead}} (\epsilon + e V) \, \nu (\epsilon) ,
\end{equation}
in terms of the densities of states $\nu_{\text{lead}} (\epsilon)$ and
$\nu (\epsilon)$ of the lead and semi-infinite wire, respectively.
Assuming $\nu_{\text{lead}} (\epsilon)$ is essentially
energy-independent, the differential conductance $d I / d V$ provides a
direct measure of the tunneling density of states at the edge of the
wire. The latter is given by
\begin{equation}\label{TDOS}
\nu (\epsilon) = \frac{1}{4 \pi (i k_F)^2} \textrm{Im} \sum_s \left.
\frac{\partial^2}{\partial x \partial x'} G^R_s (x, x', \epsilon)
\right|_{x = x' = 0} ,
\end{equation}
where $G_s^R (x,x',\epsilon)$ is the retarded Green function,
\begin{equation}\label{GR}
G_s^R (x,x',\epsilon) = - i \int_0^\infty dt \, e^{i \epsilon t}
\left\langle \left\{ \psi_s (x,t), \psi_s^\dagger (x', 0) \right\}
\right\rangle .
\end{equation}
Here, we have taken the edge of the wire at $x = 0$, and assumed
Dirichlet boundary condition, $\psi (0) = 0$. We calculate $\nu
(\epsilon)$ in the presence of a weak scatterer at $x_0$, as a
perturbation to the known result for an ideal semi-infinite Luttinger
liquid~\cite{kane-fisher}.

Following the standard Luttinger liquid derivation, we bosonize the
fermionic operators,
\begin{equation}\label{bosonization}
\psi_s (x) = \frac{1}{\sqrt{2 \pi a}} F_s \left( e^{-i k_F x + i
\Phi_{-,s} (x)} - e^{i k_F x + i \Phi_{+,s} (x)} \right) .
\end{equation}
Here, $\Phi_{+,s}$ ($\Phi_{-,s}$) are the bosonic fields corresponding
to the right (left) movers, $F_s$ is the Klein factor, and $a$ is a
short-distance cutoff. The fields $\Phi_{\pm,s}$ are decoupled
into charge ($\rho$) and spin ($\sigma$) modes, as well as into their
symmetric and anti-symmetric combinations $\vartheta$ and $\varphi$,
\begin{equation}
\Phi_{\pm,s} (x) = \frac{1}{\sqrt{2}} \{ \vartheta_\rho (x) \pm
\varphi_\rho (x) + s [\vartheta_\sigma(x) \pm \varphi_\sigma (x)] \} .
\end{equation}
At the barrier, the fields obey the boundary condition $\partial_x
\vartheta_\rho (0) = \partial_x \vartheta_\sigma (0) = 0$,
corresponding to the vanishing of the current at the origin. The beauty
of the Luttinger liquid formalism is that in terms of the bosonic
variables, the Hamiltonian becomes quadratic,
\begin{equation}\label{H0}
H_0 = \sum_{\nu = \rho,\sigma} \frac{v_\nu}{2 \pi} \int dx \left[ g_\nu
(\partial_x \vartheta_\nu)^2 + \frac{1}{g_\nu} (\partial_x
\varphi_\nu)^2 \right],
\end{equation}
where $v_\rho$ ($v_\sigma$) is the velocity of the charge (spin) mode,
and $g_\rho$, $g_\sigma$ are the Luttinger liquid parameters. Here, we
neglect the interaction in the spin channel. This simplification is
possible if the $2k_F$ component of the inter-electron interaction
potential is weak, or the energy $\epsilon$ is small, as the
interaction term is irrelevant under the renormalization flow. In
addition, in both cases $g_\sigma = 1$.

To calculate the tunneling density of states $\nu (\epsilon)$ in
perturbation theory, Eqs.~(\ref{TDOS})--(\ref{GR}) are expressed in
terms of time-ordered (and anti-time ordered) Green functions. Using
the bosonization formula, Eq.~(\ref{bosonization}), the tunneling
density of states is then expressed as
\begin{equation}\label{nu}
\nu(\epsilon) = - \frac{1}{\pi} \sum_s \mathrm{Im} \int_0^\infty dt \,
e^{i \epsilon t} \left[ \mathcal{G}_s (t) - \bar{\mathcal{G}}_s (t)
\right] .
\end{equation}
Here, $\mathcal{G}_s(t)$ is the time-ordered Green function,
\begin{eqnarray}\label{Gs}
\lefteqn{\mathcal{G}_s (t) =} \\ \nonumber & & - \frac{i}{2 \pi a}
\left\langle \hat{T} e^{i \left[ \vartheta_\rho (0, t) + s \vartheta_\sigma
(0, t) \right] / \sqrt{2}} e^{- i \left[ \vartheta_\rho (0, 0) + s
\vartheta_\sigma (0, 0) \right] / \sqrt{2}} \right\rangle ,
\end{eqnarray}
where $\hat{T}$ is the time-ordering operator. A similar expression holds for
the anti-time ordered function $\bar{\mathcal{G}}$.

In absence of the scatterer, the average in Eq.~(\ref{Gs}) is
calculated using the relation $\langle e^A e^B \rangle_0 = e^{\langle A
B + A^2 / 2 + B^2 / 2 \rangle_0}$ (valid for $A$ and $B$ linear in the
fields). Next, the correlation functions are calculated by expressing
the fields in terms of the eigenmodes of $H_0$,
\begin{equation}
\varphi_\nu (x, t) = i \sqrt{g_\nu} \! \int_0^\infty \! \frac{dp}{2
\pi} \, c_p \sin p x \left( a_{p, \nu} e^{-i v_\nu p t} \! - \!
a^\dagger_{p, \nu} e^{i v_\nu p t} \right)  \! ,
\end{equation}
\begin{equation}
\vartheta_\nu (x, t)  =  \frac{1}{\sqrt{g_\nu}} \! \int_0^\infty \frac{d
p}{2 \pi} c_p  \cos p x \left( a_{p, \nu} e^{-i v_\nu p t} \! + \!
a^\dagger_{p, \nu} e^{i v_\nu p t} \right) \! ,
\end{equation}
where $c_p = e^{- a p / 2} \sqrt{2 \pi / p}$. We then have (at $T=0$)
\begin{equation}
\mathcal{G}_s^{(0)} (t > 0) = - \frac{i}{2 \pi a} \prod_{\nu =
\rho,\sigma} \left( \frac{a}{a + i v_\nu t} \right)^{1/2 g_\nu} ,
\end{equation}
leading to the known power law dependence of the edge tunneling density
of states~\cite{kane-fisher},
\begin{equation}\label{nu0}
\nu^{(0)} (\epsilon) =  \frac{1}{\pi \Gamma \left( \frac{1}{2 g} +
\frac{1}{2} \right) \sqrt{v_\sigma v_\rho}} \left( \frac{\epsilon
a}{v_\rho} \right)^{1 / 2 g - 1 / 2} \,  .
\end{equation}
Here, we have set $g_\rho = g$ and $g_\sigma = 1$.

The leading correction to $\nu (\epsilon)$ due to an obstacle
placed at $x=x_0$, is caused by the electron backscattering from it.
The backscattering term in the Hamiltonian, which is the one relevant
to the calculation, is given by
\begin{equation}
H_1 = - \frac{U}{2 \pi a} \sum_s \left( i e^{2 i k_F x_0 + i \sqrt{2}
\left[ \varphi_\rho (x_0) + s \varphi_\sigma (x_0) \right]} +
\text{h.c.} \right),
\end{equation}
where $U$ is the $2k_F$ component of the potential created by the
scatterer. We assume $U$ is small, and treat it perturbatively. (The
opposite limit of a strong scatterer corresponds to the problem of
resonant tunneling through a double barrier, previously studied in
detail~\cite{gornyi}; this limit is not suitable for SPM measurements.)

To linear order in the scatterer strength $U$, one expands the
scattering matrix implicit in the definition of the time-ordered Green
function (both in the numerator and in the denominator).
The correction to the Green function (\ref{Gs}) is then given by
\begin{widetext}
\vspace{-0.5cm}
\begin{equation}
\mathcal{G}_s^{(1)} (t) = - \frac{1}{2 \pi a} \int_{-\infty}^\infty dt'
\left\langle \hat{T} e^{i \left[ \vartheta_\rho (0, t) + s
\vartheta_\sigma (0, t) \right] / \sqrt{2}} e^{- i \left[
\vartheta_\rho (0, 0) + s \vartheta_\sigma (0, 0) \right] / \sqrt{2}} \left[ H_1
(t') - \left\langle H_1 (t') \right\rangle_0 \right] \right\rangle_0 .
\end{equation}
Using $\langle e^A e^B e^C \rangle_0 = e^{\langle AB + AC + BC + (A^2 +
B^2 + C^2) / 2 \rangle_0}$, the first order correction to the
Green function is
\begin{eqnarray}
\lefteqn{\mathcal{G}_s^{(1)} (t > 0) = - \frac{i U}{(2 \pi a)^2}
\sum_{\pm} \pm e^{\pm 2 i k_F x_0} \left( \frac{a^2}{a^2 + 4 x_0^2}
\right)^{(g_\rho + g_\sigma) / 4}}
\\
\nonumber & & \times \int_{-\infty}^\infty dt'   \prod_{\nu = \rho,
\sigma} \left( \frac{a}{a + i v_\nu t} \right)^{1 / 2 g_\nu} \left[
\left( \frac{a - i x_0 \mathrm{sgn} (t-t') + i v_\nu |t - t'|}{a + i
x_0 \mathrm{sgn} (t-t') + i v_\nu |t - t'|} \right)^{\mp 1 / 2} \left(
\frac{a + i x_0 \mathrm{sgn} (t') + i v_\nu |t'|}{a - i x_0
\mathrm{sgn} (t') + i v_\nu |t'|} \right)^{\pm 1 / 2} - 1 \right] .
\end{eqnarray}
\end{widetext}
The calculation of the tunneling density of states, Eq.~(\ref{nu}),
thus involves integrations over $t$ and $t'$, along contours which
pass several branch cuts in the complex $t$ and $t'$ planes. Here, we
consider the contribution arising from these branch cuts assuming they
are well separated, so that the contribution of each branch cut may be
accounted for independently. The conditions for the validity of this
approximation are
\begin{equation}
\frac{\epsilon x_0}{v_\rho} \gg 1, \qquad \frac{\epsilon x_0}{v_\sigma}
\gg 1, \qquad \frac{\epsilon x_0}{v_\sigma} - \frac{\epsilon
x_0}{v_\rho} \gg 1 .
\vspace{0.2cm}
\end{equation}
The last of these three conditions leads to a separation of the
contributions of the charge and spin modes. Because of it, the results
are non-perturbative in the interaction strength, even in the limit of
weak interaction ($1 - g \ll 1$).

The contribution to the density of states induced by the scatterer is
divided in two parts. The first one comes from the branch cuts at $t' =
x_0 / v_\mu$ and $t - t' = x_0 / v_\nu$, where the velocities $v_\mu$
and $v_\nu$ each take the value of $v_\rho$ or $v_\sigma$. Setting
$g_\rho = g$ and $g_\sigma = 1$, this contribution is
\begin{eqnarray}\label{nu1}
\lefteqn{\delta \nu_1 (\epsilon) = \frac{1}{\pi^2} \,
\frac{v_\rho + v_\sigma}{v_\rho - v_\sigma} \,
\frac{U / v_\rho}{\sqrt{v_\rho v_\sigma}} \,
\frac{v_\rho}{\epsilon x_0}
\left( \frac{a}{2 x_0} \right)^{1 / 2g + g / 2 - 1} }
\\
\nonumber & & \!\!\! \times  \sum_{\mu,\nu = \rho,\sigma} C_{\mu \nu}
\cos \left[ 2 k_F x_0 + \epsilon x_0 \left( \frac{1}{v_\mu} +
\frac{1}{v_\nu} \right) + \phi_{\mu \nu} \right] ,
\end{eqnarray}
where
$C_{\rho \rho} = 1$, $C_{\sigma \sigma} = \left(
v_\sigma / v_\rho \right)^{1 / 2 g - 1/2}$,
\begin{displaymath}
C_{\rho \sigma} = C_{\sigma \rho} = \left( \frac{2 v_\sigma}{v_\rho +
v_\sigma} \right)^{1/2 g} \left( \frac{2 v_\rho}{v_\rho + v_\sigma}
\right)^{1/2} ,
\vspace{0.2cm}
\end{displaymath}
and $\phi_{\mu \nu} = (1 - 1/g) \pi / 4 + (\delta_{\mu \rho} +
\delta_{\nu \rho}) \pi / 2$. Equation~(\ref{nu1}) contains terms
oscillating with the position of the scatterer, $x_0$, with three
different wavevectors: $2 (k_F + \epsilon / v_\rho)$, $2 (k_F +
\epsilon / v_\sigma)$, and $2 k_F + \epsilon / v_\rho + \epsilon /
v_\sigma$.  Physically, these terms arise from processes involving a
right-moving excitation (either spin or charge) traveling from the edge
of the system, then scattering at $x_0$ into a left-moving excitation
(again, either spin or charge) and returning to the origin. Note the
possibility of mixing the spin and charge excitations in such
processes, leading to the oscillatory term that involves both
velocities.

The second contribution arises from the branch cuts at $t = 0$ and $t'
= \pm x_0 / v_\nu$. For this contribution, we find
\begin{eqnarray}\label{nu2}
\lefteqn{\delta \nu_2 (\epsilon) = - \frac{2^{1 / 2g - 1/2}}{\pi \Gamma
\left( \frac{1}{2 g} - \frac{1}{2} \right)} \, \frac{U /
v_\rho}{\sqrt{v_\rho v_\sigma}}}
\\
\nonumber & \times & \left( \frac{v_\rho}{\epsilon x_0} \right)^{3 / 2
- 1/ 2 g} \left( \frac{a}{2 x_0} \right)^{1 / 2g + g / 2  - 1} \cos (2
k_F x_0) .
\end{eqnarray}
The resulting oscillation is independent of energy. Motivated by the
weakly interacting limit in the spinless case, this contribution may be
understood as arising from changes to the Friedel oscillations
introduced by the scatterer. Note that $\delta \nu_2$ decays as
$x_0^{-g/2-1/2}$, the same power as the decay with position of the
Friedel oscillations in the electron density (in the absence of the
scatterer)~\cite{egger}.

To linear order in $U$, the tunneling density of states is thus a sum
of a smooth term, $\nu^{(0)}$, and oscillating terms, $\delta \nu =
\delta \nu_1 + \delta \nu_2$, given by Eqs.~(\ref{nu0}), (\ref{nu1}),
and~(\ref{nu2}). The behavior of the oscillating terms depends on the
strength of the electron-electron interaction. When $g \ll 1$,
corresponding to a strong repulsive interaction, $\delta \nu_1$ rapidly
decays with $x_0$ and $\delta \nu_2$ is the dominating term; it does
not reveal spin-charge separation. In contrast, for a weaker
interaction the oscillations in $\delta \nu_1$ and $\delta \nu_2$ decay
with $x_0$ with similar power laws (to linear order in $1-g$, we have
$\delta \nu_1 \propto x_0^{-1}$ and $\delta \nu_2 \propto x_0^{-1 + (1
- g) / 2}$). While the decay of $\delta \nu_1$ with $x_0$ is still a
bit faster than that of $\delta\nu_2$, the latter term contains a small
factor $\Gamma^{-1} (1 / 2g - 1 / 2)\approx (1 - g) / 2$. As a result,
for a broad range of $x_0$ the main contribution comes from $\delta
\nu_1$. In addition, in this weakly interacting limit, the prefactors
in Eq.~(\ref{nu1}) may be approximated by $C_{\rho \sigma} \approx
C_{\sigma \sigma} \approx C_{\rho \rho} = 1$, to obtain
\begin{eqnarray}\label{nu1approx}
\delta \nu_1 (\epsilon) & \propto & \left[ 1 + \sin \left(
\frac{\epsilon x_0}{v_\sigma} - \frac{\epsilon x_0}{v_\rho} \right)
\right]
\\
\nonumber & \times & \sin \left( 2 k_F x_0 + \frac{\epsilon
x_0}{v_\sigma} + \frac{\epsilon x_0}{v_\rho} + \phi_{\sigma \sigma}
\right) .
\end{eqnarray}
The different oscillations in $\delta \nu_1$ thus form a beating
pattern. Experimentally, this may provide a tool for the observation of
spin-charge separation.

\begin{figure}
\includegraphics[width=3.125in]{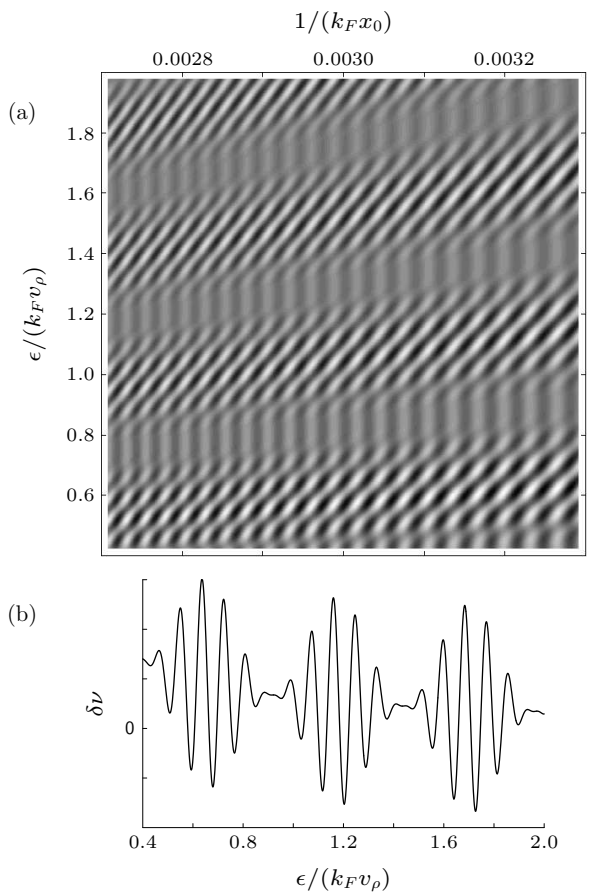}
\vspace{0.3cm} \caption{\label{fig} The oscillatory part of the density
of states, see Eqs.~(\ref{nu1})--(\ref{nu2}), in arbitrary units. The
parameters used are $g = 0.8$ and $v_\rho / v_\sigma = 1.4$. (a)
Density plot as a function of $\epsilon$ and $1 / x_0$. The main
feature of the moir\'{e} pattern, i.e., fast oscillations appearing as
slanted bright and dark regions with a slowly modulating amplitude, is
captured by Eq.~(\ref{nu1approx}).  The contribution of Eq.~(\ref{nu2})
appears as fainter vertical lines in the regions where the amplitude of
the main oscillations is small. (b) Energy dependence at $k_F x_0 =
300$.}
\end{figure}

To illustrate these ideas further, we plot $\delta \nu$ in
Fig.~\ref{fig}, using $g = 0.8$ and $v_\rho / v_\sigma = 1.4$. This
choice of parameters corresponds to a relatively weak interaction, for
which the assumptions leading to Eq.~(\ref{nu1approx}) are valid (with
these parameters, $C_{\sigma \sigma} = 0.959$ and $C_{\rho \sigma} =
0.964$).  The moir\'{e} pattern expressed by Eq.~(\ref{nu1approx}) is
clearly visible in Fig.~\ref{fig}(a).  This pattern may then be used to
obtain the spin and charge mode velocities.

A measurement of $\nu (\epsilon)$ may probe spin-charge separation even
in an experiment with a fixed (and maybe unknown) value of $x_0$.
Indeed, Eq.~(\ref{nu1approx}) shows that the number of fast
oscillations of $\delta\nu_1(\epsilon)$ fitting between two adjacent
minima of the slow amplitude modulation is $(v_\rho + v_\sigma) /
(v_\rho - v_\sigma)$.  For example, in Fig.~\ref{fig}(b) there are 6
such oscillations, corresponding to $v_\rho / v_\sigma = 1.4$.

The SPM method considered here has connections with the 
STM technique studied in Refs.~\cite{eggert1,eggert2,kivelson}. 
The differences between the two methods arise from the different 
role of the scanning tip:
In STM it is used to measure the local density of states rather than for creating 
a scattering potential for the electrons.
In both cases, the measured differential conductivity exhibits oscillations 
as a function of tip position, 
which contain information about the spin and charge velocities.
The difference in setup, however, affects the details of these oscillations and their
dependence on the Luttinger liquid parameters.
In particular, the SPM setup does not give rise to 
long wavelength oscillations of the type predicted for STM~\cite{eggert1}. 

In conclusion, the theory presented in this Letter suggests a way to study
spin-charge separation in a quantum wire by means of a scanning probe
measurement. The clearest manifestation of the separation is in the
beating pattern of the differential conductance as a function of
applied bias, see Fig.~\ref{fig}.  The SPM method may be experimentally useful
as it only requires contacting a single wire at its ends, unlike STM
or momentum conserving tunneling.
Finally, we note that our results, obtained at $T=0$, 
also hold at finite temperature,
provided the distance between the tunneling point and the position of
the probe is shorter than the thermal length $v_\sigma / T$.

We thank Rafi de Picciotto, Shahal Ilani, and Amir Yacoby for useful
discussions. This work was supported by NSF grants DMR-02-37296 and
EIA-02-10736.

\end{document}